\documentclass[12pt]{revtex4}
\usepackage{epsf}
\usepackage{graphicx}
\def\bea{\begin{eqnarray}}
\def\eea{\end{eqnarray}} 
\def\be{\begin{equation}}
\def\ee{\end{equation}}
\def\nn{\nonumber}

\def\a{\alpha}
\def\b{\beta}

\def\d{\delta}

\def\e{\epsilon}

\def\G{\Gamma}

\def\m{\mu}
\def\n{\nu}

\def\r{\rho}
\def\s{\sigma}
\def\t{\tau}
\def\th{\theta}

\begin{document}

\title {Time Evolution of Horizons} 

\vspace{1cm}

\author{ Arundhati Dasgupta}
\affiliation { Department of Physics and Astronomy, \\ University of Lethbridge, Lethbridge, T1K 3M4, Canada.}

\vspace{1cm}

\begin{abstract}
 A density matrix is defined using coherent states for space-times with apparent horizons. 
Evolving the density matrix in time gives the origin of Hawking
radiation. 

\end{abstract}

\maketitle
\section{Introduction}

A new theory is expected to take over at Planck distances as `quantum effects' of gravity start dominating.  
One of the promising approaches to the theory of quantum gravity is the theory of Loop Quantum Gravity (LQG), which is by formulation
non-perturbative and background independent \cite{lqg, lqg2, lqg3}. LQG has a well defined kinematical Hilbert space, and though the Hamiltonian constraint remains unsolved, the theory allows for a semiclassical
sector of the theory. This includes `coherent states' \cite{hall, thiem} which are peaked at classical phase space elements. Using these as a starting point,
I defined in a series of papers \cite{adg1,adg2,adg3} coherent states for the Schwarzschild space-time, and derived an origin of entropy using 
quantum mechanical definition of entropy from density matrices.
 The exact entropy is a function of the
graph used to obtain the LQG phase space variables \cite{adg4}. 
 The zeroeth order term is proportional to the area of the horizon signifying a universality of the Bekenstein-Hawking
entropy. The proportionality constant and the correction terms bring out the details of the graph \cite{adg3}. 

In this paper we take this new way of finding the origin of entropy a step further by evolving
the spatial slice in time \cite{adg5}, and observing the evolution of the density matrix in the process.
This state as of now does not satisfy the Hamiltonian constraint, but
one is allowed to take an arbitrary initial state, or a wavepacket with appropriate properties, representing a macroscopic configuration.
The evolution discussed in this paper is semiclassical, i.e. no attempt is made to use the full Hamiltonian.  

The quasilocal energy (QLE) of an outside observer,
defined in \cite{bro23}  is used as the Hamiltonian to evolve the system. 
As the time clicks in the observer's clock, the Hamiltonian evolves the coherent state such that the area of the horizon remains 
the same as predicted by classical physics. However, classically forbidden regions become accessible quantum mechanically, and 
vertices of the graph hidden behind the horizon in one slice emerge outside the horizon in the next slice. This gives a net change 
in area, and the mass deficit is emitted from the black hole. This evolution is {\it not unitary}, and the quasi-local energy 
which is used to evolve the slice is not mapped to a Hermitian operator. When matter is coupled to the gravitational system, a net 
flux emerges causing a decay of the horizon.

In section II we introduce the formalism by describing the coherent state, the black hole time slice, the apparent horizon equation, 
and the density matrix. Section III describes the time
evolution of the system and gives a derivation of the change in entropy. In section IV we give a description of a matter current
emergent from behind the horizon. Finally in the concluding section we include a discussion about the implications of the non-Hermitian evolution. 

\section{The coherent state in LQG}
For gravity, finding the canonical variables which describe the physical phase space is an odd task
as there is no unique time. Nevertheless a fiducial time coordinate can be chosen, which breaks the manifest 
diffeomorphism invariance, restored in the Hilbert space of states by imposing constraints. 

The constant time slices are described by the intrinsic metric $q_{ab}$ and the extrinsic curvature $K_{ab}$ (a,b=1,2,3). The theory can be formulated in terms of the square root of the metric, the triads $e^I_a$ defined thus:
\be
e^I_a e^I_b= q_{ab}
\ee
where $I$ represents the internal index for the rotation group SO(3) of the tangent space and $a,b=1,2,3$. The internal group is taken to be SU(2), as this is locally isomorphic to SO(3).
The theory is then defined in terms of the `spin connection' $\G^I_a = \e^{IJK}e^b_J\nabla_a e_{bK}$ and the triads. However, 
a redefinition of the variables in terms of tangent space densitised triads $E_a^I$
and a corresponding gauge connection $A_a^I$ where $I$ represents the SU(2)
index simplifies the quantisation considerably. 
\be
A_a^I =\G_a^I - \beta K_{ab} e^{I b} \ \ \ \ E^a_I = \frac1{\beta} ({\rm det} \ e) e^a_I
\label{defn} 
\ee
($e_a^I$ are the usual triads, $K_{ab}$ is the extrinsic curvature, $\G_a^I$ the associated spin connection, $\beta$ the one parameter ambiguity which remains named as the Immirzi
parameter.)
The quantisation of the Poisson algebra of these variables is done by smearing the connection along one dimensional edges $e$ of length $\delta_e$ of a graph $\G$ to get holonomies $h_e(A)$. The triads are smeared in a set of
2-surface decomposition of the three dimensional spatial slice to get the corresponding momentum $P_e^I$. The algebra is then represented in a kinematic `Hilbert space', in which the physical constraints
have been `formally' realised \cite{thiemdiet}. Once the phase space variables have been identified, one can write
a coherent state for these \cite{hall} i.e. minimum uncertainty states peaked at classical values of $h_e,P^I_e$. In analogy with the harmonic oscillator coherent states,
where the coherent state is a function of the complexified phase space element $x-ip$, the SU(2) coherent states are peaked at the complexified phase space
element $g_e= e^{i T^I P^I/2} h_e$. These $g_e$ are thus elements in the complexification of SU(2) as $e^{i T^I P^I/2}$  ($T^I$ being the generator matrices of SU(2)) is a Hermitian matrix and $h_e$ is the unitary SU(2) matrix. 
Whether these are physical coherent states, or have appropriate behavior under the action of the constraints has to
be examined carefully \cite{thiem2}. The coherent state in the momentum representation for one edge is defined to be
\be
|\psi^t(g_e)>= \sum_{jmn} e^{-t j(j+1)/2} \pi_j (g_e)|jmn>
\label{coherent}
\ee
In the above $g_e$ is a complexified
classical phase space element $e^{i T^I P^{I \rm cl}_e/2} \ h^{\rm cl}_e$, (the $P^{I \rm cl}_e$ and the $h^{\rm cl}_e$ represent classical momenta and holonomy
obtained by embedding the edge in the classical metric). The $|jmn>$ are the usual basis spin network states given by $\pi_j (h)_{mn}$, which is the jth representation of the SU(2) element $h_e$.
 Similarly, $(2j+1) \times (2j+1) $ dimensional representations of the 2$\times2$ matrix $g_e$ are denoted as $\pi_j(g_e)_{mn}$. The j is the
quantum number of the SU(2) Casimir operator in that representation, and $m,n$ represent azimuthal quantum numbers which run from $-j..j$.
  The coherent state is precisely peaked with maximum probability at the $h^{cl}_e$ for the
variable $h_e$ as well as the classical momentum $P_e^{I \rm cl}$ for the variable $P_e^{I}$. The fluctuations about the classical value are controlled by the parameter $t$ (the semiclassicality parameter). This parameter is given by $l_p^2/a$ where $l_p$ is Planck's constant and $a$ a dimensional constant
which characterises the system.
The coherent state for an entire slice can be obtained by taking the tensor product of the coherent state for each edge which form a graph $\G$,
\be
\Psi_{\G}= \prod_e \psi^t_e .
\label{coh}
\ee
 In \cite{adg1} the $g_e$ was evaluated for the 
Schwarzschild black hole by embedding a graph on a spatial slice with zero intrinsic curvature. The particular graph which was used had the edges along the coordinate lines of a sphere. This simplistic graph, was very useful in
obtaining the description of the space-time in terms of discretised holonomy and momenta. A particularly interesting consequence of this was that
the phase space variables were finite and well defined even at the singularity. 

Given that the area of a surface in gravity is measured as the integral of the square root of the metric
over the surface, the area operator can be written simply as $\hat A= \sqrt{\hat P^I_e \hat P^I_e}$.
The expectation value of the area operator in the coherent state emerges as \cite{adg4}
\be
<\psi|\hat A|\psi> = (j+ \frac12) t
\ee

{\it Thus we are considering a semiclassical state, which is a state such that expectation values of operators are closest
to their classical values. The information of the classical phase space variables are encoded in the
complexified SU(2) elements labeled as $g_e$. The fluctuations over the classical values are controlled by the semiclassical
parameter $t$.} 

The density matrix which describes the entire black hole slice is obtained as
\be
\rho^{\rm Total}= |\Psi_\G><\Psi_\G|
\ee
where $|\Psi_\G>$ is the coherent state wavefunction for the entire slice, a tensor product of coherent state for each edge.

\subsection{Apparent Horizons}
We concentrate on the coherent state near the apparent horizon contained in the spatial slice.
 We find that motivated from the apparent horizon equation the graph across the horizon can be taken to be populated
by radial edges, linking vertices outside and inside the horizon. One then traces over the coherent state
within the horizon.
Initially we take a particular time
slicing of the black hole, which has the spatial slices with zero intrinsic curvature \cite{adg1}.  
One such metric which has the time slices as flat is the Lemaitre metric
\be
ds^2 = -d\tau^2 + \frac{dR^2}{\left[\frac{3}{2r_g}(R-\t)\right]^{2/3}} + \left[\frac{3}{2}(R-\t)\right]^{4/3} r_g^{2/3}(d\theta^2 +\sin^2\theta d\phi^2).
\ee
The $r_g= 2GM$, (in units of c=1)and
in the $\tau=$ constant slices one can define the induced metric in terms of a $`r'$ coordinate defined as $dr =dR/\left[3/2r_g (R- \tau_c)\right]^{1/3}$ ($\tau=\tau_c$) on the slice. One gets the metric of the three slice to be   
\be
ds_3^2= dr^2 + r^2 (d\theta^2 + \sin^2\theta d\phi^2).
\label{sph}
\ee
The entire curvature of the space-time metric is contained in the extrinsic curvature or $K_{\m\n}=\frac12 \partial_\tau g_{\m \n}$ tensor of the $\tau=$ constant slices. 
Now if there exists an apparent horizon somewhere in the above spatial slice, then that is located as
a solution to the equation
\be
\nabla_a S^a + K_{ab}S^aS^b -K=0
\ee
where $S^a$ , ($(a,b=1,2,3)$ denote the spatial indices) is the normal to the horizon, $K_{ab}$ the extrinsic curvature in the induced coordinates of the slice, and $K$ the trace of the extrinsic
curvature. If the horizon is chosen to be the 2-sphere, then in the coordinates 
of (\ref{sph}), $S^a\equiv(1,0,0)$,
the apparent horizon equation as a function of the metric reduces to:
\be
K_{rr}(1-q^{rr}) -K_{\phi\phi}q^{\phi \phi} - \G^{\phi}_{\phi r} -K_{\th\th}q^{\th \th}-\G^{\th}_{\th r}=0
\label{classical}
\ee

Note that the first term of the equation disappears trivially as $1=q^{rr}$ for any point in the spatial slice. 
Even at the operator level the $q^{rr}$ can be set to the identity operator in the first approximation, as
$q^{rr}= P_{e_r} P_{e_r}/{V}^2$ ($\hat V$ being the volume operator) upto normalisations, and in the spherically symmetric 
metric $ V= P_{e_r} \delta_{e_r}$ (upto discretisation constants). Thus the operators in the numerator and denominator cancel and the normalisation 
conspire, leaving $\hat q^{rr}=\cal I$. 
To understand the rest of the equation in terms of the holonomy and momentum variables of LQG, which are classically
measured in the same metric as (\ref{sph}), we use the following regularisation

\be
K_{\th(\phi),\th(\phi)}= e^I_{\th(\phi)}K^I_{\th(\phi)}, \ \  q_{\th(\phi)\th(\phi)}=e^I_{\th(\phi)}e^I_{\th(\phi)}
\label{ext}
\ee

\be
e^I_{\th,(\phi)} \equiv N\ {\rm  Tr}[T^I h^{-1}_{e_{\th(\phi)}}\{h_{e_{\th(\phi)}},V\}]
\label{triads}
\ee

(N is a constant, a function of the edge lengths and the area bits of the discretisation) and
V is the volume operator.

\be
K^I_{\th(\phi)}=\frac{1}{\d_{e_\theta}}{\rm Tr}[h_{e_\theta}^{-1} T^I \beta\frac{\partial}{\partial\beta}h_{e_\th}]
\label{ext2}
\ee
 Here $\beta$ has been used as a parameter to identify the $K^I_a$ operator, and this is mainly a trick. 
 In the continuum limit
\be
h_{e} (A^I_a) = {\rm Limit}_{\d_{e_a} \rightarrow 0} e^{\int {\bf A_a} dx^a}= (I + A^I_a T^I \delta_{e_a})
\ee
As the gauge connection is a function of the Immirzi parameter due to (\ref{defn}), the expectation value of this operator in 
a coherent state will be a function of the Immirzi parameter.
By taking the derivative wrt to the Immirzi parameter
we are giving the same status to the parameter as is given to `dimension' in a dimensional regularisation of Feynman diagrams.
We let the parameter vary by an infinitesimal amount from its value in the particular quantisation sector, take the
derivative, and put its original value in the final answer for the $K_a^I$ operator. The formula (\ref{ext2}) is facilitated
by the fact that the dependence of $A_a^I$ on the $\beta$ is linear. One way to check whether this gives the proper answer
is to take a solved quantum mechanical system and use a similar method there. The most useful example is the Harmonic Oscillator
Hamiltonian, which can be written as
\be
H= \frac{p^2}{2m} + \frac{1}{2} m \ \omega^2 x^2 
\ee

The ground state is a coherent state, so we take that as an example. We define the operator
\be
x^2 = \frac{2}{m} \frac{\partial H}{\partial \omega^2}
\ee
Thus
\be
<x^2> = \frac{2}{m} <\frac{\partial H}{\partial \omega^2}>= \frac{2}{m}\frac{\partial}{\partial \omega^2} \left(\frac{\hbar \omega}{2}\right)= \frac{\hbar}{2m\omega}
\ee
The regularisation (\ref{ext2}) is thus an allowed approximation.

The terms involving the Christoffel connections like $\G_{\th r}^{\th}$ include
derivatives in the regularised version, the derivatives appear as difference of triads across two vertices. Thus
\be
\G^{\theta}_{\theta r}= e^{\theta}_I e^{\theta}_I\frac{1}{\delta_{e_r}}\left(e^J_\theta (v_1) - e^J_{\theta}(v_2)\right)e^J_{\theta}(v_1)
\ee
As a result of this if we impose restrictions on the Christoffel connections
 and one of the vertices $v_1$ is within the horizon, whereas $v_2$ is outside the horizon, there will be correlations
 across the horizon. 

If one evaluates the expectation value of the apparent horizon equation using the regularised variables in the coherent states, then one 
would obtain 
\bea
&&4<\psi|P_{e_\theta}^2 \left[{\rm Tr}\left(T^J h_{e_{\theta}}^{-1} V^{1/2} h_{e_{\theta}}\right)_{v_1}- {\rm Tr}\left(T^J h_{e_\th}^{-1}V^{1/2}h_{e_\th}\right)_{v_2}\right]{\rm Tr}\left(T^J h_{e_\th}^{-1}V^{1/2}h_{e_\th}\right)_{v_2}|\psi> \nn \\
&&- N' <\psi| {\rm Tr}\left(h_{e}^{-1} T^I \beta \frac{\partial}{\partial \beta} \ h_{e_\th}\right) P^{I}_{e_\th}\ _{v_1}|\psi> =0 
\label{diff12}
\eea
($N'$ is a constant)

\subsection{Density Matrix}

The density matrix is obtained as
\be
\rho^{\rm Total}= |\Psi_\G><\Psi_\G|
\ee
where $|\Psi_\G>$ is the coherent state wavefunction for the entire slice, a tensor product of coherent state for each edge.

But given this, we concentrate in a `local' region to see the behavior of the horizon 

\be
\rho^{\rm Total}= \rho^{\rm outside}\rho^{\rm local}\rho^{\rm inside}
\ee
where $\rho^{\rm local}$ covers a band of vertices surrounding the horizon one set on a sphere at radius $r_g - \delta_{e_r}/2$ and one set
on a sphere at radius $r_g + \delta_{e_r}/2$ within the horizon, as described in \cite{adg4}, and in the figure enclosed.
This local density matrix and the correlations due to the apparent horizon equation (\ref{diff12}) was used
to derive entropy \cite{adg2}. This entropy counts the number of ways to induce the horizon area
using the spin networks, though the constraints have not been appropriately imposed as was obtained
using a Chern-Simons theory in \cite{abck}. However, the entropy calculation using the coherent states 
provides a tracing mechanism, and a method to obtain correlations across the horizon which are gravitational in origin.
We will henceforth deal with $\rho^{\rm local}$, but we will drop the ${\rm local}$ label for brevity.

\begin{center}
\includegraphics[scale=0.3]{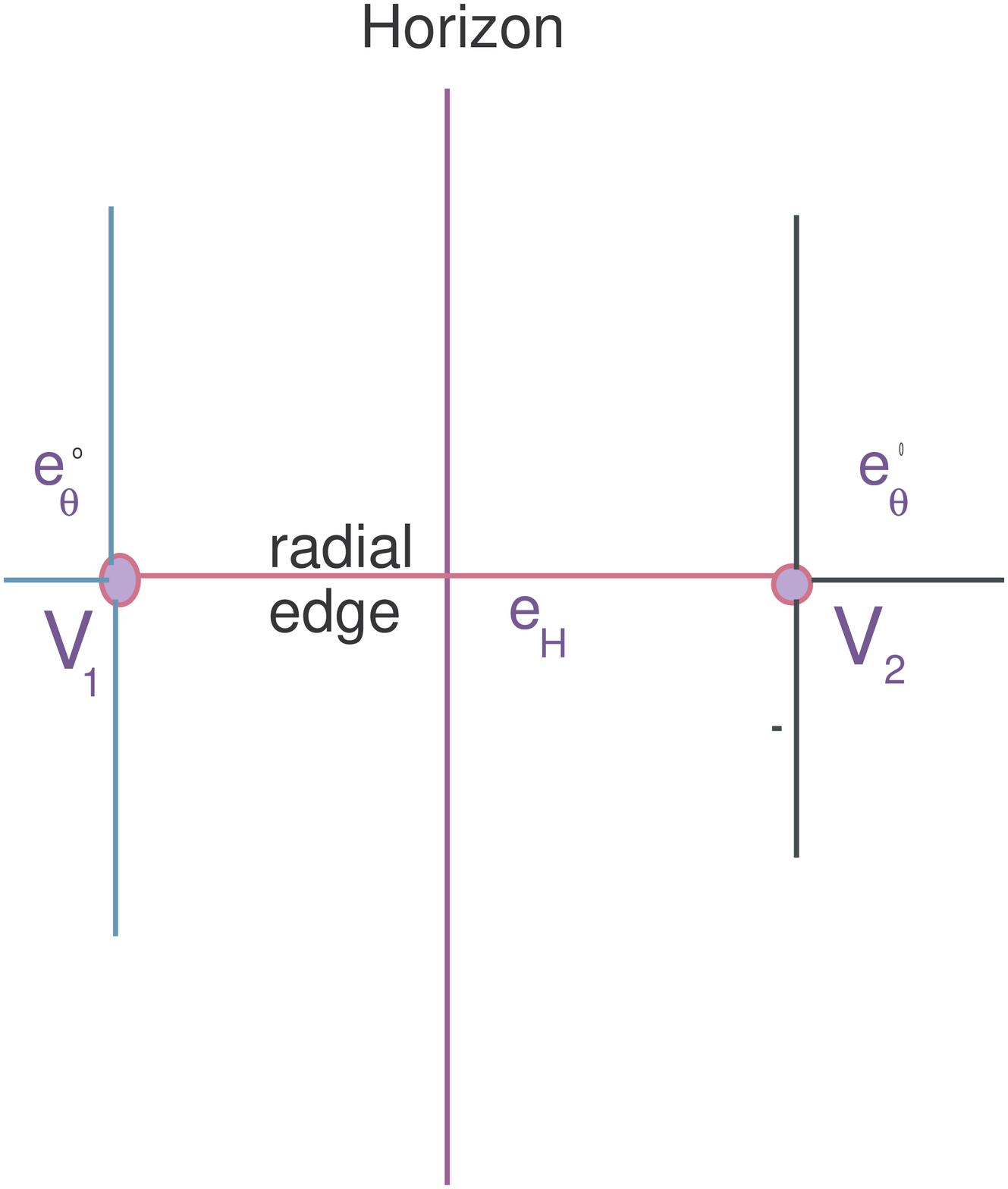}
\end{center}
\section{Time Evolution}
In physical systems, the Hamiltonian generates time evolution, but in General Theory of Relativity, 
 the Hamiltonian is a constraint and generates diffeomorphisms in the time direction. So the question is, what is physical time, and if that exists,
what would be the operator evolving the system in that direction? In case of space-times
with time like Killing vectors, notion of time can be identified with the Killing direction, and a notion of `quasilocal energy' (QLE) defined using
the same. The QLE then generates translations in the Killing time. In case of the Schwarzschild space-time,
the QLE has been defined in \cite{bro23}. We build the Hamiltonian
which evolves the horizon from one time slice to the next by appropriately regularising the QLE.
Note the `Killing time' and QLE are classical concepts, and thus regularising QLE gives us a `semiclassical' Hamiltonian.

\subsection{Change in Entropy}

Before we get into the analysis of what QLE evolution means,
we take a simple system made up of two subsystems, and examine the consequences of a Hamiltonian evolution.
Let the density matrix be defined for a system whose states are given in the tensor product Hilbert space $H_1 \otimes H_2$
and given by
\be
|\psi>= \sum_{ij} d_{i j} |i>|j>
\ee
where $|i>$ is the basis in $H_1$ and $|j>$ is the basis in $H_2$ and $d_{ij}$ are the non-factorisable coefficients of the wavefunction
in this basis.  
Let us label the wavefunction at time $t=0$ to be given by the coefficients $d^{0}_{ij}$.
The density matrix is
\be
\rho^{0}= \sum_{ ij i' j'} d^{0 *}_{i' j'} d^{0}_{i j} |i>|j><j'|<i'|
\ee

The reduced density matrix if one traces over $H_2$ is:

\be
{\rm Tr_2}\rho^{0}= \sum_{i i'}\sum_j d^{0 *}_{i' j} d^0_{i j}|i><i'|
\ee

We now evolve the system using a Hamiltonian which has the matrix elements $H_{i j i' j'}|i>|j><j'|<i'|$, we 
assume that the Hamiltonian does not factorise, that is there exists interaction terms between the two Hilbert spaces. 
The evolution equation is:
\be
i \hbar \frac{\partial \rho}{\partial \t}= [ H, \rho]
\ee

which in this particular basis gives the density matrix elements at a infinitesimally nearby slice to be
\be
d^{\delta\t *}_{i' j'} d^{\delta \t}_{i j} = d^{0 *}_{i' j'} d^{0}_{i j} - \frac{i}{\hbar} \delta \t \left[\sum_{k l} \left(H_{ij kl} d^{0}_{k l} d^{0 *}_{i' j'} - d^{0}_{i j} d^{0 *}_{kl} H_{kl i' j'}\right)\right]
\ee
Thus we evolve the `unreduced' density matrix and then trace over the $H_2$ in the evolved slice.
The reduced density matrix in the evolved slice is:
\be
\sum_j d^{\delta\t *}_{i' j} d^{\delta \t}_{i j} = \sum_j d^{0 *}_{i' j} d^{0}_{i j} - \frac{i}{\hbar} \delta \t \left[\sum_{k l j } \left(H_{ij kl} d^{0}_{k l} d^{0 *}_{i' j} - d^{0}_{i j} d^{0 *}_{kl} H_{kl i' j}\right)\right].
\ee

This gives:

\be
\rho^{\delta \t} = \rho^{0} -\frac{i}{\hbar} \delta \t A
\ee

where $A$ represents the commutator. Clearly the entropy in the evolved slice evaluated as $S_{\rm BH}^{\delta \t}= -Tr(\rho \ln \rho)$ can be found as
\be
S_{\rm BH}^{\delta \t} = S_{\rm BH}^{0} +\frac{i}{\hbar} \delta \t [{\rm Tr} A \ln \rho^0 + {\rm Tr}\rho^{0}\rho^{0 \ -1} A]
\label{change}
\ee

Given the definition of $A_{ii}$, one gets
\be
A_{ii} = \sum_{jkl} \left[\rho^0_{ijkl}H_{klij}- H_{ijkl}\rho^0_{klij}\right]
\ee
In case both the Hamiltonian and the density operator are Hermitian, one obtains
\be
\sum_j A_{jj}=  2 \iota \ {\rm Im Tr}(\rho^0 H)
\label{eqn}
\ee
 
This is clearly calculable, and gives the change in entropy $\Delta S_{\rm BH}$. The $\ln \rho^0$ term yields corrections, and we ignore it in the first approximation.

\subsection{The Hamiltonian}

To trace the origin of Horizon fluctuations, we must take an observer who is stationed
outside the horizon, or in other words is not a freely falling observer. 
The quasilocal energy is defined using a `surface' integral of the extrinsic curvature with which the surface is embedded in three space. In
our case, we take the bounding surface to be the horizon and the quasilocal energy is given by
the surface term\cite{haw23,bro23}.
\be
\tilde H= \frac{1}{\kappa}\int d^2x \sqrt{\s} k
\ee
where $k$ is the extrinsic curvature with which the 2-surface, which in this case is the horizon $S^2$ is embedded in the spatial 3-slices,
and $\s$ is the determinant of the two metric $\s_{\m \n}$ defined on the 2-surface. This `quasilocal energy' is measured with reference
to a background metric. Thus $H=\tilde H -H_o$.  We concentrate on the physics observed in a observer stationed at a $r=$ constant
sphere.
 
 The metric in static $r=const$ observer's frame is 
\be
ds^2= - f^2 dt^2 + r^2 (d\th^2 + \sin^2\th d\phi^2).
\ee
The $f= \sqrt{1- r_g/r}$ where $r_g$ is the Schwarzschild radius. 
If we take $n_{\m}$ to be the space-like vector,
normal to the 2-surface, then the extrinsic curvature is given by:
\be
k_{\m \n}= \s_{\m}^{\a}\nabla_{\a}n_{\n}
\ee
and the trace is obviously
\be
k= \nabla^{\a}n_{\a}
\ee
In the special slicing of the 
of the stationary observer the normal to the horizon 2-surface is given by $(0, f(r) , 0,0)$.
However, we built the coherent state on the Lemaitre slice. The Lemaitre and the Schwarzschild observer's coordinates are related by the following coordinate transformations,

\be
\sqrt{\frac{r}{r_g}} dr  = (dR - d\tau) \ \ \ 
dt = \frac{1}{1-f'}\left(d\tau - f' dR \right)\ \ \ f'  = \frac{r_g}{r}
\label{tran1}
\ee

The r= const cylinder of the Schwarzschild coordinate corresponds to $dR= d\tau$ of the Lemaitre coordinates,
and for these $dt=d\tau$. Thus unit translation in the $t$ coordinate coincides with unit translation in the
$\tau$ coordinate. Further, the intersection of the $r=$ constant cylinder with a $t=$ constant surface
coincides with the intersection of r=constant and the $\tau$=constant surface. 
Thus in the initial slice, the QLE Hamiltonian can be written as
\be
H= \frac1{2 \kappa}\int d\theta d\phi \sqrt{g_{\th \th} g_{\phi \phi}} [-g^{\th \th}\frac{\partial g_{\th \th}}{\partial r} - g^{\phi \phi} \frac{\partial g_{\phi \phi}}{\partial r}]f(r) - H_0
\label{hamil1}
\ee
The reference frames' quasilocal energy is a number, it just defines the zero point Hamiltonian. Thus, we replace the classical expressions by operators evaluated at the $\tau=$ constant slice. In the first approximation 
we simply take the $f(r)$ as classical $\sqrt{1-r_g/r}=\sqrt{\delta_{e_r}/2r_g}=\epsilon$, as this arises due to the coordinate transformation and the norm of the vector $n_r$ in the previous frame. 
In the re-writing of (\ref{hamil1}) in regularised LQG variables the Hamiltonian
appears rather complicated.  

One can rewrite these in a much simpler form,
using the apparent horizon equation. Since the Hamiltonian is an integral over the horizon, the variables will satisfy the apparent horizon Equation (\ref{classical}) upto quantum fluctuations. Thus the Hamiltonian operator is then re-written as
\be
H_{\rm horizon}= \frac \e{\kappa}\int d\theta d\phi \sqrt{g_{\th \th} g_{\phi \phi}} [K^I_{\theta}e^{I\theta} + K^I_{\phi} e^{I \phi}]
\ee
where we have used the classical apparent horizon equation (\ref{classical}) (with $q^{rr}=1$).

\be
H_{\rm Horizon} = \frac{C a \e}{2\kappa \delta_{e_\th}s_{e_\th}} \sum_{v_1}{\rm Tr}[h_{e_\th}^{-1} T^I \beta\frac{\partial}{\partial \beta} h_{e_\theta}] P_{e_\theta I} + h.c. + (\th\rightarrow\phi)
\label{hamil2}
\ee
where $C$ consists of some dimensionless constants $s_{e_\th}$ is the 2-dimensional area bit over which $E^{\theta}_I$ is smeared, $a$ is a dimensionfull constant which appears to get the $P_{\ e_\theta I}$ dimension less. $\delta_{e_{\th}}$ is the length for the
angular edge $e_{\th}$ over which the gauge connection is integrated to obtain the holonomy. The sum over $v_1$ is the set of vertices immediately
outside the horizon. The (\ref{hamil2}) can then be lifted to an operator.

{\it This regularised expression for QLE is for the horizon 2-surface only and would not apply for any other spherical surface in the Schwarzschild space-time.}

\subsection{ U(1) Case}

Let us take the U(1) case to make the calculations easier and observe the action of the QLE Hamiltonian on the evolution of the coherent state. The spin network states are replaced by $|n>= e^{\iota n\zeta}$ , $0<\zeta<2\pi$, n is an integer and the 
coherent states are:
\be
\psi^t(g_e)= \sum_n e^{-(t n^2)/2} e^{i n(\chi_e - ip_e)} e^{-\iota n \zeta}
\ee

$g_{n  \ e}= e^{i n (\chi_e -i p_e)}$ is the complexified phase space element in the `nth' representation.

The QLE operator also takes the simplified form 
\be
H_{\rm Horizon}^{U(1)}= -\frac12 C'\iota \hat h_{e}^{-1}\beta \frac{\partial}{\partial \beta} \hat h_e \hat p_e + \frac12 C'\iota \hat p_e  \beta \frac{\partial \hat h_e^{-1}}{\partial \beta} \hat h_e
\label{hamil}
\ee

The prefactors have been clubbed into $C'$.

In the calculation of the matrix elements, we drop the label of the edges $e$ for the Hamiltonian.
\be
<m|\hat H_{\rm Horizon}^{U(1)}|n> = \int e^{-\iota m \zeta} H_{\rm Horizon}^{U(1)} e^{\iota n \zeta} d\zeta
\label{adg}
\ee
This calculation can be done by putting 
an assumption that the $\zeta=\zeta_1 + \beta\zeta_2$. In this $\zeta_1, \zeta_2$ are completely independent of $\beta$.
It is an allowed assumption, and identifies the $\beta$ dependence of the operator matrix elements, which are otherwise `hidden'. 
The calculation however introduces an arbitrariness
in the formula, which can be fixed by requiring that the expectation value of the Hamiltonian
agrees with the classical QLE \cite{adg5}. However, in this paper
we use the `annihilation' operators defined in \cite{thiemwinkler}.

This is done by observing that the U(1) coherent states are eigenstates of an annihilation operator defined thus:
\be
\hat g_e = e^{t/2} e^{\hat p_e} \ \hat h_e \ \ \ \ \hat g_e |\psi> = g_e |\psi>
\ee

The holonomy operator can thus be written as
\be
\hat h_e = e^{-t/2} e^{-\hat p_e} \ \hat g_e
\ee
And the derivative wrt Immirzi parameter of the holonomy which appears in the definition of the Hamiltonian replaced by 
\bea
\beta\frac{\partial \hat h_e}{\partial \beta} &= & e^{-t/2} \left[ -\beta \frac{\partial \hat p}{\partial \beta} e^{- \hat p_e} \ \hat g_e + e^{-\hat p_e} \ \beta\frac{\partial \hat g_e}{\partial \beta}\right]\\
&=& e^{-t/2} \left[ \hat p_e \ e^{-p_e} \ \hat g_e + e^{-\hat p}\  \beta\frac{\partial \hat g_e }{\partial \beta} \right]
\eea

The dependence of the operator $p$ on the Immirzi parameter is known (\ref{defn}), and thus we could evaluate the
derivative ($\beta \partial_{\beta} \  p_e (\beta) = \beta\partial_{\beta} (p_e(1)/\beta)= -  p_e (\beta)$)
 
The term
\be
{\rm Tr}(\rho^0 H^{\rm U(1)}_{\rm Horizon})
\label{tr}
\ee

is then computable. Let us take the first term of (\ref{hamil}) and find (\ref{tr}). As $\rho^{0}= |\psi><\psi|$, (\ref{tr}) gives simply (we drop the `e' label for brevity)
\bea
<\psi|H^{\rm U(1)}_{\rm Horizon} |\psi> &=& <\psi|-\frac12 C'\iota \hat h^{-1}\beta \frac{\partial}{\partial \beta} \hat h \ \hat p|\psi> + <\psi|h.c.|\psi> \\
  &=&-\frac12 \iota C' <\psi| \hat g^{\dag} e^{-t/2} e^{-\hat p} e^{-t/2} \left[ \hat p \ e^{-\hat p} \hat g + e^{-\hat p} \b \frac{\partial \hat g}{\partial \beta} \right]\hat p|\psi> + <\psi|h.c. |\psi> \nn \\
&=&-\frac12 \iota C' e^{-t} g^* \left[<\psi|e^{-\hat p}\ \hat p \ e^{-\hat p}\hat p \ |\psi> g + <\psi| e^{-2 \hat p} \ \b\frac{\partial \hat g}{\partial \beta}\hat p \ |\psi>\right] + <\psi| h.c. |\psi> \nn 
\eea

We then concentrate on the 2nd term of the above

\bea
&& <\psi|e^{- 2\hat p}\b\frac{\partial \hat g}{\partial \beta}\hat p |\psi>\\
&=& <\psi|e^{-2 \hat p}\b\frac{\partial \hat g}{\partial \beta} \int d\nu(g') |\psi'><\psi'| \hat p |\psi> \\
&=& \int d \nu (g') \b \frac{\partial g'}{\partial \beta} <\psi| e^{-2 \hat p}|\psi'><\psi'|\hat p |\psi> \\
\eea
Where we have used the fact that coherent states resolve unity. It can be shown that the expectation value of the
operators in the $t \rightarrow 0$ collapses the integral to $g'=g$ point \cite{thiemwinkler}. Thus one obtains from the above

\bea
{\rm Tr}(\rho^0 H^{U(1)}_{\rm Horizon}) &= & - \frac12 \iota C' e^{-t/2} \left[ p + g^* e^{-2 p} \ \b\frac{\partial g}{\partial \beta}\right]p + h.c. \\
&=& C' \b \frac{\partial \chi}{\partial \beta} p
\eea
which is real, and thus 
\be
\Delta S_{\rm BH}=0
\ee

(this is actually the classical QLE as it should be from ${\rm Tr} (\rho^0 H_{\rm Horizon})$).

This is obvious, as the way the Hamiltonian is defined, this is simply a function of the Hilbert space
outside the horizon, and the matrix elements of this will not yield anything new.
 We approximated the horizon sphere by summing over
$v_1$ vertices immediately outside the horizon. We could do the same by summing over $v_2$
vertices immediately within the horizon. For the Lemaitre slice, the metric is smooth at the
horizon, and one can take the `quantum operators' evaluated at the vertex $v_2$.
In this case however, as the region is within the classical horizon, the norm of the Killing vector
is negative, and $n_r$ has components which are imaginary. The  $\epsilon\rightarrow \pm \iota \epsilon$.
Thus $H_{\rm Horizon}= \frac12 [\sum_{v_1}H_{v_1} + \sum_{v_2} H_{v_2}]$.
In the evaluation of the QLE, the energy would emerge correct in the $ \d_{e_r}\rightarrow 0$ limit as $\e\rightarrow 0$
The regularised Hamiltonian is not Hermitian, and the evolution equation is
\be
\iota \hbar \frac{\partial \rho}{\partial \tau} =  H \rho -  \rho H^{\dag}
\ee
And thus the operator which appears in the change of entropy equation is
\be
\Delta S_{\rm BH} = \frac{\iota \delta \tau}{\hbar} {\rm Tr}[H \rho^0 - \rho^0 H^{\dag}]
\ee 

\be
\Delta S_{BH}=   \mp \frac{\delta \tau}{ \hbar} C' \beta\frac{\partial \chi}{\partial \beta} \  p
\ee
The `rate of change' of entropy is thus
\be
\dot \Delta S_{BH}= \mp \frac{\tilde C}{l_p^2} \beta \frac{\partial \chi}{\partial \beta} \ p
\ee 
(we extracted the $\kappa$ from $C'$ to get $l_p^2$ and rewrote the rest of the constants as $\tilde C$). 

Thus there is a net change in entropy,
but, to see if this is Hawking radiation, we have to couple matter to the system. 

\subsection{SU(2) Case}

The SU(2) case is easily reduced to the U(1) case in the actual calculation due to the gauge fixing.
This is achieved by making the following observations: To retain the metric as in the same form as the classical metric,
we impose the conditions at the operator level
\be
P_{e_a}. P_{e_b}=0
\ee
such that the corresponding metric has only the diagonal terms as non-zero. With these additional `constraints' on the
operators, we can put the $P^I_{e_a}$ such that each has only one component surviving, let's say $P_{e_{\theta}}^I= \delta^I_3 P_{e_{\theta}}$. This also makes the holonomy
restricted to the U(1) case, as the gauge connection $A^I_{{e_\th}}$ gets restricted to the $I=3$ and other directions can be put to zero. Thus we can take the holonomy
to be diagonal
\be
h_e= \left(\begin{array}{cc}e^{\iota\zeta}&0\\0&e^{-\iota\zeta}\end{array}\right)
\ee
The operator is then obtained as
\be
H={\rm Tr}[h_{e}^{-1} T^I \beta \frac{\partial h_e}{\partial \beta}]P_e^I= {\rm Tr}[h_e^{-1}T^3 \beta \frac{\partial h_e}{\partial\beta}]P_e^3= \beta\frac{\partial \zeta}{\partial \beta}P_e^3
\ee
This is same as the U(1) Hamiltonian (upto normalisations). The spin network states also project on to U(1) subgroup, thus giving us the same techniques to use in the
calculation of the U(1) states as for this one. To observe this,
the non-zero elements for the holonomy matrix 
\be
h=\left(\begin{array}{cc}a&b\\-\bar b & \bar a \end{array}\right)
\ee
in the jth representation is given by:
\be
\pi_j(h)_{mn} = \sum_l \frac{\sqrt{(j-m)!(j+m)!(j-n)!(j+n)!}}{(j-m-l)!(j-n-l)!(m-n-l)!l!} a^{j-n+l} \bar a^{j-m+l}b^{m-n-l}\bar b^{l}
\ee
Clearly in the particular case we are considering, the $b=0$, and m,n=-j and j.
Thus the two non-zero elements are
\be
\pi_j(h)_{jj}= e^{2j \ \zeta} \ \ \ \ \ \pi_j(h)_{-j -j}= e^{- 2j \ \zeta}
\ee
The sum over j in the ${\rm Tr}(\rho^0 H_{\rm Horizon})$ with the coherent state defined in (\ref{coherent}) thus reduces to the U(1) case in the computation of the change in entropy.
Thus the rate of change in entropy of a classically spherically symmetric black hole is given by
\be
\Delta S_{\rm BH}= \mp\frac{\tilde C \delta \tau}{l_p^2} \sum_{v} \beta \left[\frac{\partial \chi_{e_\th}}{\partial \beta} P_{e_{\th}} + \frac{\partial \chi_{e_\phi}}{\partial \beta} P_{e_{\phi}}\right]
\ee

where the classical holonomies $h_{e_{\th(\phi)}}= e^{\iota \chi_{e_{\th(\phi)}}}$. If we plug in the actual values, we get this to be
\be
\Delta S_{\rm BH} = \pm \frac{2 C \epsilon \delta \t}{l_p^2} \sum_{v} dA_v 
 \beta r_g 
 \ee
where $dA_v$ the area element at vertex $v$ on the sphere.
This change in entropy is totally gravitational in origin, and seems to signify the emergence of `geometry' from within the horizon. 

In fact, if we some over the area, we get the $\Delta S_{\rm BH} = \pm \frac{8 \pi \e \delta \t} {l_p^2} r_g $ (if we set $C=1/\beta$),
which would be the change in entropy when the radius of the horizon changes by $\delta r_g = \epsilon \delta \t$!

\section{Outgoing flux of radiation}
In the previous section we found that as the system evolves in time, the horizon fluctuates and the area decreases. But is this Hawking radiation? Adding matter to a `coherent state' description
of semiclassical gravity has been discussed \cite{thiemsahl}. Thus, given a massless scalar field Lagrangian coupled to gravity, whose Hamiltonian
is given by
\be
H_{\rm sc}= \int d^3x  \ [\frac{ \pi^2}{\sqrt{q}} + (\nabla \phi)^2 ],
\ee
the `gravity' in the Hamiltonian can be regularised in terms of the $h_e, P^I_e$ operators in the coherent state formalism. The integral over the three volume
gets converted to a sum over the vertices dotting the region. Thus
\be
H_{\rm sc}^{v} = \sum_{v} H_{v} (h_e, P^I_e, V)
\ee

This Hamiltonian is an operator, and one evaluates
an expectation value of the Hamiltonian in the reduced density matrix of the initial slice, to find the classical behavior of the scalar field as observed by an observer
outside the horizon. Thus
\be
{\rm Tr}\left(\rho^{\tau} H^{\tau}_{\rm sc}\right)
\ee

This Hamiltonian and the density matrix are then both evolved according to the time-like observers frame. One gets 
\be
i \hbar \frac{\partial H_{\rm sc}}{\partial \tau}= [H, H_{\rm sc}]
\ee
This gives
\be
{\rm Tr}\left(\rho^{\tau+ \delta \tau} H^{\tau + \delta \tau}_{\rm sc}\right) - {\rm Tr} \left(\rho^{\t} H_{\rm sc}^{\t}\right)= -(\d \t)^2 {\rm Tr} \{[H,\r^{\t}] [H, H_{\rm sc}^{\t}]\}
\label{fluct}
\ee

It is very clear thus that the order $\d \t$ terms are zero for this. However, allowing for the non-unitary evolution using the non-Hermitian
Hamiltonian, the $\d \t$ terms survive. In fact the terms are
\be
-\frac{\iota \d \t}{\hbar}{\rm Tr}[(H \rho - \rho H^{\dag})H_{\rm sc}] - \frac{\iota \d \t}{\hbar}{\rm Tr}[\rho (H H_{\rm sc} - H_{\rm sc} H^{\dag})]
\ee
The first term is remarkable, it shows that the term giving rise to entropy change teams up with the expectation value of the scalar Hamiltonian. The second
term yields corrections, so we ignore it in the first approximation.
The exact details of the computation have to be obtained using the coherent state of the matter and gravity coupled system \cite{thiemsahl}. If one
simple takes the matter + gravity system in a tensor product form, and one has matter quanta of energy $\omega$  
sitting at one vertex, then the first term would give new matter in the evolved slice as $\Delta S_{\rm BH}\ \omega$. The `rate' of particle creation thus has the form $- 2\epsilon\omega/T_H$ where
$ T_H$ is the Hawking temperature for the signs $+ (-)\iota \epsilon$ and negative (positive) $\omega$.
 
Thus from the above it seems\\
(i)One has found emission of matter quanta from a black hole but from a `semiclassical' description rooted in a theory of quantum gravity, beyond quantum fields in curved space-time.\\
(ii)The results indicate a non-unitary evolution which allows space to emerge from within the horizon.\\
(iii) The emission is perceived by a static or an accelerating observer as anticipated, and the non-unitary flow might be due to the semiclassical
approximations. A quantum evolution using the quantum Hamiltonian might still be unitary.

The above derivation seems to be a `quantum gravity' description of the tunneling mechanism for describing Hawking radiation \cite{tunnel}. However, 
the results are preliminary and further investigation has to be done. 

\section{Conclusion}
In this paper we showed a method to obtain the origin of Hawking radiation using a coherent state description of a black hole
space-time. We took a quantum wavefunction defined on an initial slice, peaked with maximum probability at classical phase space-variables.
 We then evolved the slice using a Hamiltonian, which is the 
`quasilocal energy' at the horizon. This QLE evolved the system in the time 
 and the entropy was shown to change, indicating a change in black hole mass and hence an emergence of
interesting non-unitary dynamics. One of course has to investigate further to see 
 what is the endpoint of this time evolution. The time flow indicates one might have to formulate quantum theory of gravity rooted in irreversible physics. 
The presence of additional degrees of freedom in the form of `graphs' also indicates that the classical phase space might not be described
by deterministic physics, but by distributions, a manifestation of microscopic irreversible physics in complex
systems. 

{\bf Acknowledgements} This research is supported by NSERC; research funds of University of Lethbridge. I would like to thank
B. Dittrich for useful discussion; J. Supina for proofreading the manuscript.

\end{document}